\documentclass[aoas,nameyear,dvips]{arximspdf}
\usepackage{dcolumn}
\usepackage{stfloats}
\usepackage{graphicx}

\doi{10.1214/09-AOAS310}
\volume{4}
\issue{1}
\pubyear{2010}
\firstpage{151}
\lastpage{178}

\makeatletter
\fnbelowfloat
\newcolumntype{d}[1]{D{.}{.}{#1}}
\renewcommand{\epsilon}{\varepsilon}
\renewcommand{\cite}[1]{\citet{#1}}

\newcommand{\bb}{{\mathbf b}}
\newcommand{\be}{{\mathbf e}}
\newcommand{\bx}{{\mathbf x}}
\newcommand{\bW}{{\mathbf W}}

\newcommand{\bphi}{{\bolds{\phi}}}

\makeatother

\begin{document}
\begin{frontmatter}

\title{E-loyalty networks in online auctions}
\runtitle{E-loyalty networks in online auctions}

\begin{aug}
\author[A]{\fnms{Wolfgang} \snm{Jank}\ead[label=e1]{wjank@rhsmith.umd.edu}\corref{}}
\and
\author[A]{\fnms{Inbal} \snm{Yahav}\ead[label=e2]{iyahav@rhsmith.umd.edu}}
\runauthor{W. Jank and I. Yahav}
\affiliation{RH Smith School of Business and University of Maryland}
\address[A]{RH Smith School of Business\\ University of Maryland\\
4370 VMH\\
College Part 20742\\
USA\\
\printead{e1}\\
\phantom{E-mail: }\printead*{e2}} 
\end{aug}

\received{\smonth{12} \syear{2008}}
\revised{\smonth{11} \syear{2009}}

%
\begin{abstract}
Creating a loyal customer base is one of the most important, and at
the same time, most difficult tasks a company faces. Creating
loyalty online (e-loyalty) is especially difficult since customers
can ``switch'' to a competitor with the click of a mouse. In this
paper we investigate e-loyalty in online auctions. Using a unique
data set of over 30,000 auctions from one of the main
consumer-to-consumer online auction houses, we propose a novel
measure of e-loyalty via the associated network of transactions
between bidders and sellers. Using a bipartite network of bidder and
seller nodes, two nodes are linked when a bidder purchases from a
seller and the number of repeat-purchases determines the strength of
that link. We employ ideas from functional principal component
analysis to derive, from this network, the loyalty distribution
which measures the perceived loyalty of every individual seller, and
associated loyalty scores which summarize this distribution in a
parsimonious way. We then investigate the effect of loyalty on the
outcome of an auction. In doing so, we are confronted with several
statistical challenges in that standard statistical models lead to a
misrepresentation of the data and a violation of the model
assumptions. The reason is that loyalty networks result in an
extreme clustering of the data, with few high-volume sellers
accounting for most of the individual transactions. We investigate
several remedies to the clustering problem and conclude that loyalty
networks consist of very distinct segments that can best be
understood individually.
\end{abstract}

%

\end{frontmatter}

\section{Introduction}

Online auctions are becoming an increasingly important component of
consumers' shopping experience. On eBay, for instance, several
million items are offered for sale every day. What makes online
auctions popular forms of commerce is their availability of almost
any kind of item, whether it be new or used, and their constant
accessibility at any time of the day, from any geographical region
in the world. Moreover, the auction mechanism often engages
participants in a competitive environment and can result in
advantages for both the buyers and the sellers
[\citet{bajari2004eii}].

In this paper we study online auctions from the point of view of the
bidder--seller network that they induce. Every time a bidder
purchases from a seller, both bidder and seller are linked. Buying
from a seller indicates that the bidder likes the product and trusts
the seller---thus, it establishes a relationship between bidder and
seller. Many sellers list more than one auction (i.e., they sell
multiple items across different auctions), so repeat transactions by
the same bidder across different auctions of the same seller measure
the strength of this relationship, that is, it measures the strength
of a bidder's \textit{loyalty} to a particular seller.

Studying loyalty in auction networks is new. Much of the existing
auction literature focuses on only the seller and the level of \textit{trust}
she signals to the bidders
[e.g., \citet{brown2006roa}]. To
that end, a seller's \textit{feedback score} (i.e., the number of
positive ratings minus the number of negative ratings) is often
scrutinized [e.g., \cite{Lingfang06}] and it has been shown that
higher feedback scores can lead to price-premiums for the seller
[see \cite{Reiley00}; \cite{livingston2005vgr}]. In this paper we study a
complementary determinant of a bidder's decision process: \textit{loyalty}.
Loyalty is different from trust. Trust is often associated
with reliability or honesty; and trust may be a necessary (but not
sufficient) prerequisite for loyalty. Loyalty, however, is a
stronger determinant of a bidder's decision process than trust.
Loyalty refers to a state of being faithful or committed. Loyalty
incorporates not only the level of confidence in the outcome of the
transaction, but also satisfaction with the product, the price, and
also with previous transactions by the same seller. Moreover, loyal
bidders are often willing to make an emotional investment or even a
small sacrifice to strengthen a relationship. This paper makes two
contributions to the literature on online auctions: First, we
propose a novel way to \textit{measure} e-loyalty from the bipartite
network of bidders and sellers; then, we investigate the \textit{effect}
of e-loyalty on the outcome of an auction and the
statistical challenges associated with it.

More specifically, our goal is to understand and learn from loyalty
networks. To that end, we first measure a seller's perceived loyalty
by its induced bidder loyalty distribution. Then, borrowing ideas
from \textit{functional data analysis}, we capture key elements of that
distribution using functional principal component analysis. The
resulting principal component scores capture different aspects of
loyalty-strength, -skew and -variability. We then investigate the
impact of these loyalty scores on the outcome of an auction such as
its final price.

We would like to point out that the goal of this paper is not to
develop new auction theory (i.e., it is not our goal to develop a
game-theoretic model under market equilibrium considerations).
Rather, our goal is to mine a rich set of auction data for new
patterns and knowledge. In that sense, our work is exploratory
rather than confirmatory. However, as it is often the case with
exploratory work, we hope that our work will also inspire the
development of new theory. In particular, we hope that our work will
bring the attention to the many statistical challenges associated
with the study of online markets.

Studying e-loyalty networks is challenging from a statistical point
of view because of the asymmetric nature of the network. Just as in
many offline markets, online auctions are dominated by few very
large sellers (``Megasellers''). Megasellers have a large supply of
products and thus account for a large number of all the
transactions. Statistically, this dominance results in a clustering
of the data and, as a result, a violation of standard OLS model
assumptions. In this paper we investigate several remedies to this
clustering effect via random effects models and weighted least
squares. However, our investigation shows that neither approach
fully eases all problems. We thus conclude that the data is too
segmented to be captured by a single model and compare our analyses
with the results of a data-clustering approach.

This paper has implications for future research in online markets.
Many online markets are characterized by a few large ``players'' that
dominate most of the interactions and many, many small players with
occasional interactions. This is often referred to as the ``long
tail effect'' in online markets [see, e.g., \cite{bailey2008}]. For
instance, on eBay, Megasellers dominate the marketplace. The
statistical implication is that repeat interactions by these
Megasellers are no longer independent and, hence, the assumptions of
OLS break down. While this may not always create a problem, this
research shows that, first, the conclusions from an OLS regression
are significantly different from models that account for the
clustering induced by Megasellers, and, second, that it is not at
all obvious how to best account for this clustering. In particular,
this research puts the spotlight on the findings from previous
researchers [e.g., \cite{Reiley00}; \cite{bapavlou2002}; \cite{bapnaisr2008}]
who, despite similar data-scenarios, rely their conclusions on the
OLS modeling assumptions. [See also \citet{bajari2004eii} who, in
the context of trust and online auctions alone, count over 6 papers
relying on OLS modeling techniques.]

This paper is organized as follows. In Section~\ref{sec2} we introduce our
data and we motivate the existence of auction networks. In
Section~\ref{sec3}
we use seller--bidder networks to derive several key measures of
e-loyalty. We investigate the effect of e-loyalty on the outcome of
an auction in Section~\ref{sec4} and explore different modeling alternatives.
The paper concludes with final remarks in Section~\ref{sec5}.

\section{Auction networks}\label{sec2}
In this section we describe the data for our study. We start by
describing the online auction mechanism in general and our data in
particular. Then, we motivate the network structure induced by the
auction mechanism and show several snapshots of our network data.

\subsection{Online auction data}

In online auctions participants bid for products or services over
the Internet. While there are different types of auction mechanisms,
one of the most popular types (a variant of which can also be found
on \href{http://eBay.com}{eBay.com}) is the \textit{Vickrey} auction, in which the
initial price starts low and is bid up successively. Online auctions
have experienced a tremendous popularity recently, which can be
attributed to several features: Since the auction happens online, it
is not bound by any temporal or geographical constraints, in stark
contrast to its brick-and-mortar counterpart (e.g., at \textit{Sotheby's}).
It also fosters social interactions since it engages
participants in a competition. As a result, it attracts a large
number of buyers and sellers which offers advantages for both sides:
sellers find a large number of potential customers which often
results in higher prices and lower costs. On the other hand, buyers
find a large variety of products which enables them to locate rare
products and to choose between products with the lowest price. One
of the most well-known online auctions is eBay, but there are many
more (e.g., uBid, Prosper, Swoopo or Overstock), each of which
offers a variety of different products and services. The data for
this research originates from eBay online auctions and we describe
details of the data next.

\begin{table}
\caption{Auctions and products}\label{tab:tab 1}%
\begin{tabular}{@{}ld{3.2}d{2.2}d{3.2}@{}}
\hline
\textbf{Attribute} & \multicolumn{1}{c}{\textbf{Mean}} & \multicolumn{1}{c}{\textbf{Median}} &
\multicolumn{1}{c@{}}{\textbf{St. dev}} \\
\hline
Auction duration (Days) & 3.50 & 3.00 & 4.43 \\
Starting price (USD) & 3.77 & 3.33 & 5.64 \\
Closing price (USD) & 6.61 & 4.25 & 9.15 \\
Item quantity & 5.42 & 1.00 & 129.47 \\
Bid count & 3.16 & 1.00 & 4.26 \\
Size (bead diameter) & 6.41 & 6.00 & 3.35 \\
Pieces (\# of beads per item) & 124.30 & 48.00 & 343.79 \\
\hline
\end{tabular}
\end{table}

\begin{table}[b]
\caption{Sellers}\label{tab:tab 2}
\begin{tabular*}{220pt}{@{\extracolsep{\fill}}ld{4.2}d{3.2}d{6.2}@{}}
\hline
\textbf{Attribute} & \multicolumn{1}{c}{\textbf{Mean}} & \multicolumn{1}{c}{\textbf{Median}} &
\multicolumn{1}{c@{}}{\textbf{St. dev}} \\
\hline
Volume & 163.90 & 6.00 & 999.00 \\
Conversion rate & 0.67 & 0.67 & 0.33 \\
Seller feedback & 2054.00 & 264.00 & 12\mbox{,}400.00 \\
\hline
\end{tabular*}
\end{table}

We study the complete bidding records of \textit{Swarovski} fine beads
for every single auction that was listed on eBay between April,
2007, and January, 2008. (Note that the data were obtained directly
from eBay so we have a complete set of bidding records for that time
frame.) Our data contains a total of 36,728 auctions out of which
25,314 transacted. There are 365 unique sellers and 40,084 bidders
out of which 19,462 made more than a single purchase. Each bidding
record contains information on the auction format, the seller, the
bidder, as well as on product details. Tables \ref{tab:tab 1}--\ref{tab:tab 3} summarize this information.

Table \ref{tab:tab 1} shows information about the auctions and the
product sold in each auction. We can see that the typical
auction-length is 3 days.\footnote{We excluded fixed-price listings
(``Buy-It-Now'') since these do not constitute true auction
mechanisms.} The product sold in each auction (packages of beads
for crafts and artisanship) is of relatively small value and, thus,
both the average starting and closing prices are low. While many
eBay auctions sell only one item at a time (e.g., laptop or
automobile auctions), auctions in the crafts category often feature
multi-unit auctions, that is, the seller offers multiple counts of
the same item and bidders can decide how many of these items they
wish to purchase. In our data the average item-quantity per auction
is 5.42. Auctions thrive under competition among bidders and while
the average number of bids is slightly larger than 3, the median is
only 1. As pointed out above, the items sold in these auctions are
packages of Swarovski beads. The value of a bead is, in part,
defined by its size, and the average diameter of our beads equals
6.41~millimeters. Another measure for the value of an item is the
number of pieces per package; we can see that there are on average
over 124 beads in each package, but this number varies significantly
from auction to auction.


\begin{table}
\caption{Buyers}\label{tab:tab 3}
\begin{tabular*}{220pt}{@{\extracolsep{\fill}}ld{3.2}d{2.2}d{3.2}@{}}
\hline
\textbf{Attribute} & \multicolumn{1}{c}{\textbf{Mean}} & \multicolumn{1}{c}{\textbf{Median}} &
\multicolumn{1}{c@{}}{\textbf{St. dev}} \\
\hline
Volume & 3.62 & 1.00 & 14.29 \\
Item quantity & 5.05 & 2.00 & 29.25 \\
Bidder feedback & 228.10 & 70.00 & 559.53 \\
\hline
\end{tabular*}
\end{table}

We are primarily interested in the bipartite network between bidders
and sellers. One main factor influencing this network is the size of
the seller. We can see (Table~\ref{tab:tab 2}) that the average
seller-volume (i.e., number of auctions per seller) is over 163. A
seller's auction will only transact if (at least one) bidder places
a bid. While low transaction rates (or ``conversion rates'') are a
problem for many eBay categories (e.g., automobiles), in our data the
average conversion rate is 67\% per seller, which is considerably
high. One factor driving conversion rates is a seller's perceived
level of trust. Trust is often measured using a seller's feedback
rating computed as the sum of positive (``$+$'') and negative (``$-$'')
ratings. Trust averages over 2000 in our data.

Table \ref{tab:tab 3} shows the corresponding attributes of the
bidders. Bidders win on average almost 4 auctions (``volume'') and,
in every auction, they purchase on average over 5 items. (Recall the
multi-unit auctions with several items per listing.) The bidder
feedback [computed as the sum of positive (``$+$'') and negative
(``$-$'') feedback] captures a bidder's experience with the
auction-process and its average is over 220 in our data, signaling
highly experienced bidders.

\subsection{Bidder--seller networks}

Interactions in an online auction result in a network linking its
participants. Bidders bidding on one auction are linked to other
bidders who bid on the same auction. Sellers selling a certain
product are linked to other sellers selling the same product. In
this study we focus on the network \textit{between} buyers and sellers.
Each time a bidder transacts with a particular seller, both are
linked.\footnote{Note that in our data, bidders and sellers form
disjoint groups, that is, a node is either a bidder-node or a
seller-node, but not both. Thus, our network forms a \textit{bipartite
network}.} A seller can set up more than one auction, thus, repeat
transactions (i.e., purchases) measure the strength of this link. For
instance, a bidder transacting 10 times with the same seller has a
stronger link compared to a bidder who transacts only twice. In our
analyses, we only consider edges with link-strength of at least 4.
That is, we disregard all bidder--seller transactions with frequency
less than 4. While there exists no recommended or ideal cut-off, our
investigations suggest that results vary for smaller values but
stabilize for link-strengths of 4 and higher. In that sense, the
network strength measures an important aspect about the relationship
between buyers and sellers: customer loyalty.

We would like to emphasize that one can measure loyalty in different
ways. While one could count all the repeat \textit{bids} a bidder
places on auctions hosted by the same seller, we only count the
number of \textit{winning bids} (i.e., the number of transactions).
While both bids and winning bids indicate a relationship between
buyers and sellers, a winning bid signals a much stronger commitment
and is thus much more indicative of a buyer's loyalty.

In this paper we investigate loyalty relationships across auctions.
Studying cross-auction relationships is rather rare in the
literature on online auctions, and it has gained momentum only
recently [\cite{Haruvy2008new}; \cite{ReddyDass2006}; \cite{jankshmueli2007}; \cite{jankzhang2008}]. In this work we consider network effects between
auction participants and their impact on the outcome of an auction.

\begin{figure}

\includegraphics{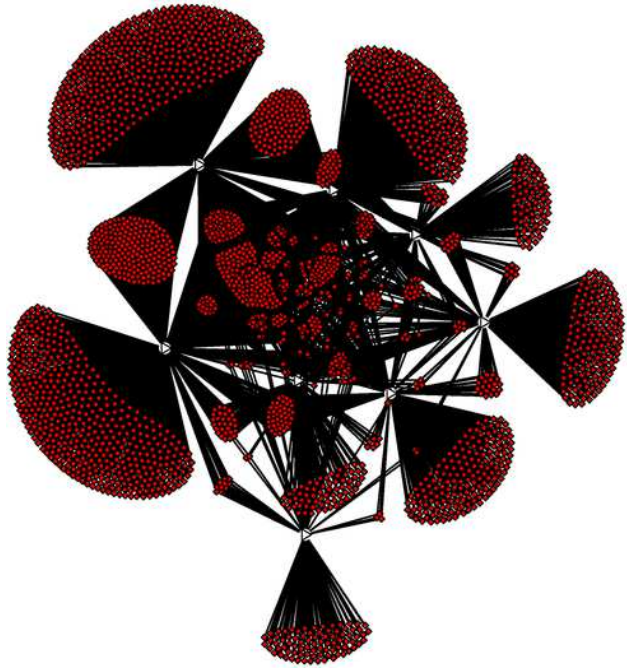}

\caption{Network of the top 10th percentile of sellers: the top
10\% of sellers hosting the most auctions.}
\label{fig:net 2}
\end{figure}

Consider Figure \ref{fig:net 2} which shows the top 10\% of high
volume \textit{sellers}. Sellers are marked by white triangles, bidders
are marked by red squares. A (black) line between a seller and
bidder denotes a transaction. The width of the line is proportional
to the number of transactions and hence measures the strength of a
link. We can see that some sellers interact with several hundred
different bidders (with 895, on average); we can also see that some
sellers are ``exclusive'' in the sense that they are the only ones
that transact with a set of bidders (see, e.g., at the margins of the
network), while other sellers ``share'' are a common set of bidders.
Serving bidders exclusively vs.~sharing them with other sellers has
huge implications on the outcome of the auction.

\begin{figure}

\includegraphics{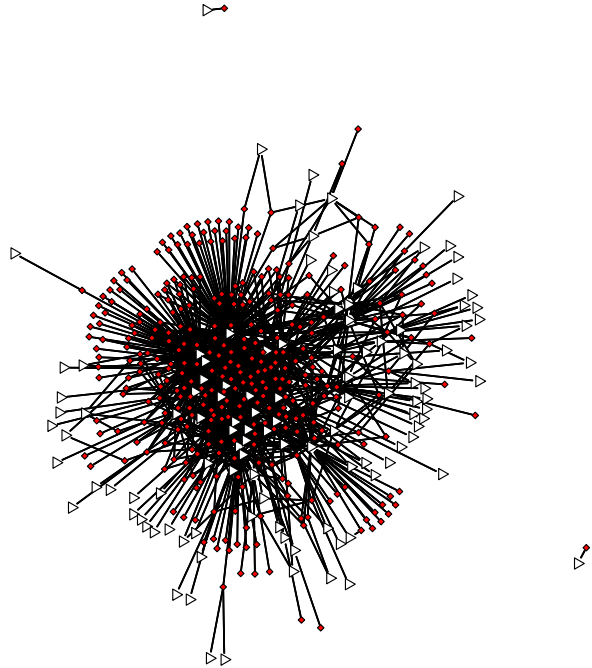}

\caption{Network of the top 10th percentile of bidders:
the top 10\% of bidders bidding on most auctions. }
\label{fig:net 4}
\end{figure}

\begin{figure}

\includegraphics{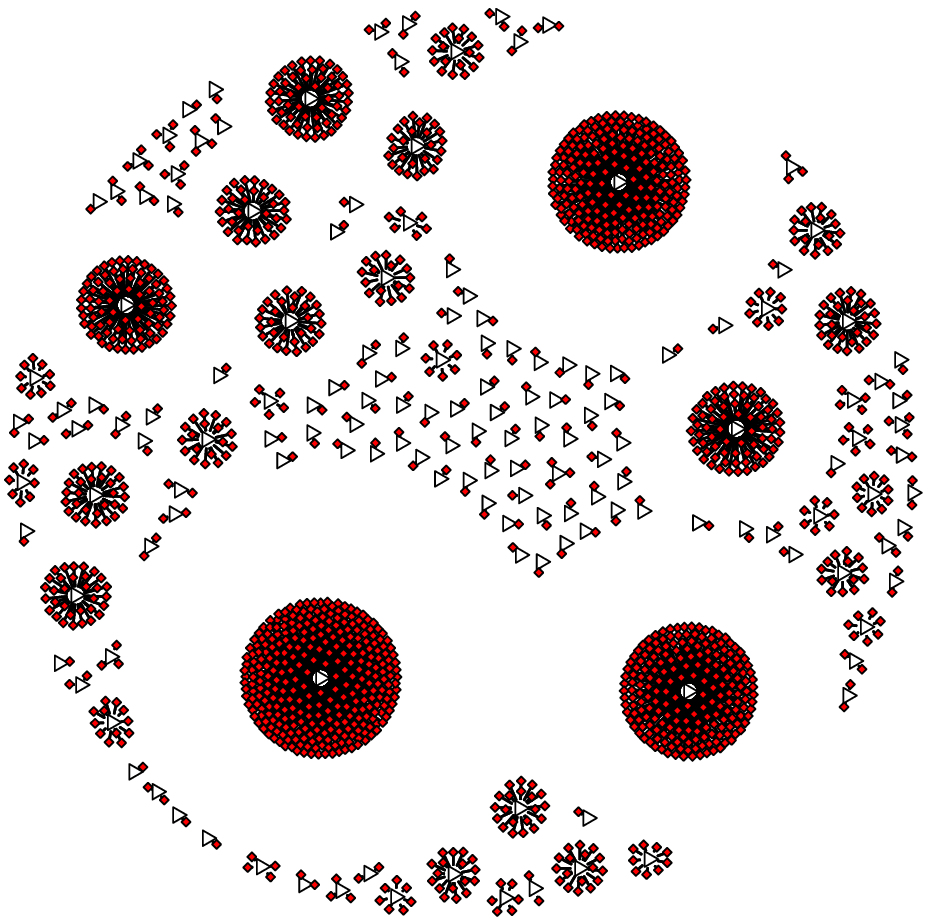}

\caption{Network of new bidders: Bidders who bid for the very first time.}
\label{fig:net 3}\vspace*{-4pt}
\end{figure}

Figure \ref{fig:net 4} shows another subset of the data. In this
figure we display only the top 10\% of all \textit{bidders} with the
highest number of transactions. We can see that many of these
high-volume bidders transact with only one seller (note the many the
red triangles which are connected with only a single arc to the
network) and are hence very loyal to the same person. Figure
\ref{fig:net 3} shows only new buyers (i.e., bidders who won an
auction for the first time). This network exemplifies the market
share of a seller with the effect of repeat buyers removed. We can
see that the market is dominated by few mega sellers, yet smaller
sellers still attract some of the buyers. We can identify 5
mega-sellers, 3 high-volume sellers, and many medium- and low-volume
sellers. Since these are only first-time buyers, loyalty does not
yet play a role in bidders' decisions. However, the fact that most
first-time bidders ``converge'' to only a few mega-sellers suggests
that this is a very difficult market for low-volume sellers to
enter.

As pointed out above, bidder--seller networks capture loyalty of
participants. While most sellers and bidders are linked to one
another, here we only focus on the sub-graphs created by each
bidder--seller pair. Next, we describe an innovative way to extract
\textit{loyalty measures} from these graphs.

\section{Extracting loyalty from network information}\label{sec3}

Our loyalty measures map the entire network of bidders and sellers
into a few seller-specific numbers. For each seller, these numbers
capture both the \textit{proportion of bidders} loyal to that seller,
as well as the \textit{degree of loyalty} of each bidder. We derive the
measure in two steps. First, we derive, for each seller, the \textit{loyalty distribution};
then, we summarize that distribution in a few
numbers using functional principal component analysis. We describe
each step in detail below.

Note that there exists more than one way for extracting loyalty
information from network data. We chose the route of loyalty
distributions since they capture the two most important elements of
loyalty: the proportion of customers loyal to one's business, and
the degree of their loyalty. Notice, in particular, that we do not try
to dichotomize loyalty (i.e., categorize it into loyal vs.~disloyal
buyers): Since we do not believe that loyalty can be turned on or
off arbitrarily, we allow it to range on a continuous scale between
0 and 1. This will allow us to quantify the impact of the \textit{shape} of a
seller's loyalty distribution on his or her bottom line.
For instance, it will allow us to answer whether sellers with \textit{pure}
loyalty (i.e., all buyers 100\% loyal) are better off compared
to sellers with more variation among their customer base.

We would also like to caution that the resulting analysis is complex
since we first have to characterize the infinite-dimensional loyalty
distributions in a finite way, and subsequently interpret the
resulting characterizations. The resulting interpretations are more
complex than, say, employing user-defined measures of loyalty (e.g.,
summary statistics such as the number of loyal buyers or the proportion
of at least 70\% loyal buyers). While such user-defined measures are
easy to interpret, there is no guarantee that they capture all of
the relevant information. (For instance, measuring the ``number of
loyal bidders'' would first require us to define a cutoff at which
we consider one buyer to be loyal and another one to be disloyal---any such
cutoff is necessarily arbitrary and would lead to a
dichotomization which we are trying to avoid.) Rather than employing
arbitrary, user-defined measures, we set out to let the data speak
freely and first look for ways to summarize the information captured
in the loyalty distributions in the most exhaustive way. This will
lead us to the notion of principal component loyalty scores and
their interpretations. We will elaborate on both aspects below.

\subsection{From loyalty networks to loyalty distributions}

\begin{figure}
\begin{tabular}{c}
\footnotesize{(a)}\\

\includegraphics{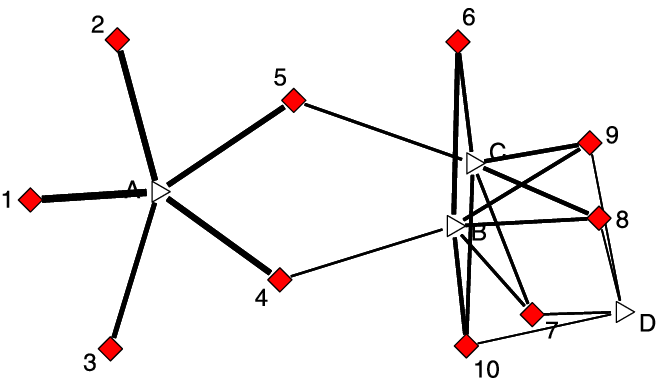}
\\
\footnotesize{(b)}\\
{
\includegraphics{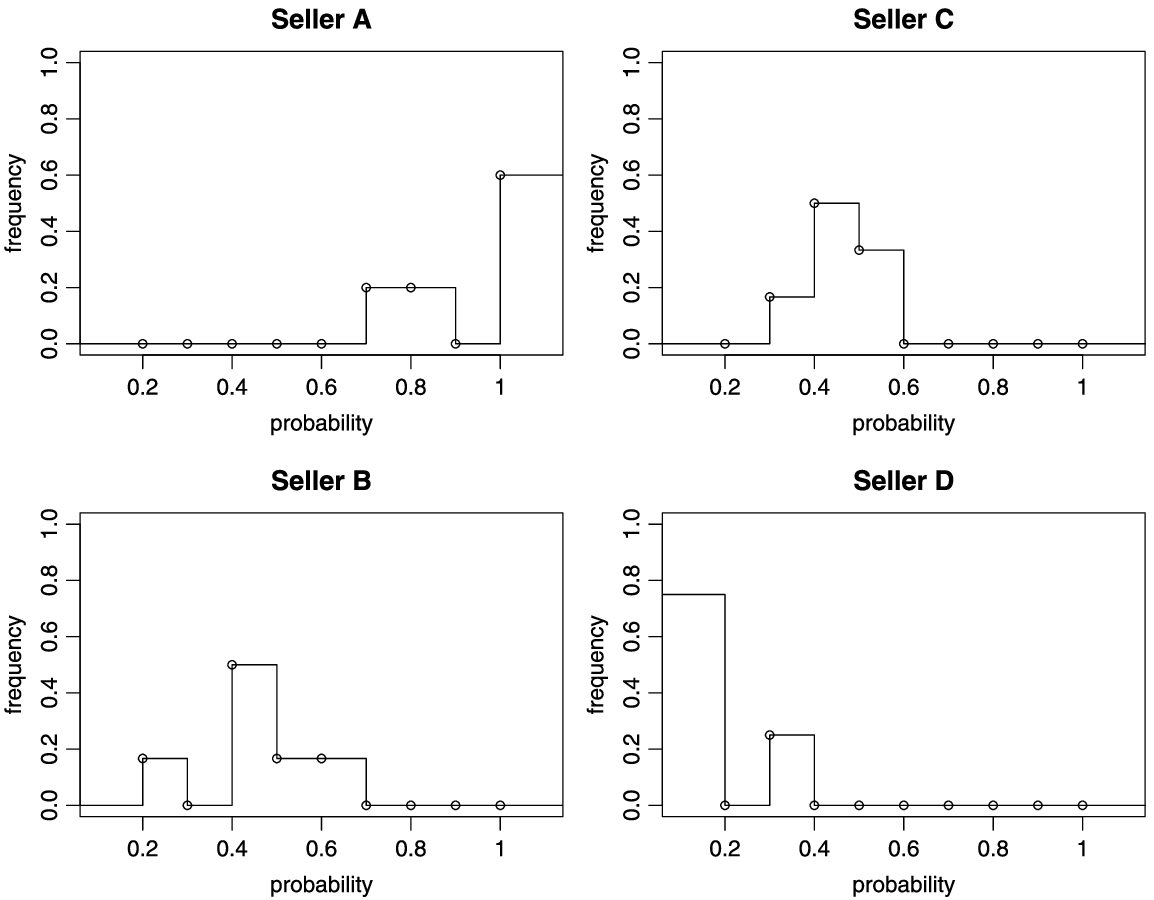}
}
\end{tabular}
\caption{\textup{(a)} Shows a hypothetical network
between~4 sellers (\textup{A--D}, white triangles) and~10 bidders
(1--10, red squares). The size of the arc between a seller
and a bidder corresponds to the number of interactions
between the two. \textup{(b)} Shows the resulting
loyalty distributions for each seller (i.e., it shows the
relative frequencies).}
\label{fig:illustration}
\end{figure}

Consider the hypothetical seller--bidder network in Figure
\ref{fig:illustration}. In that network, we have 4 sellers (labeled
``A,'' ``B,'' ``C'' and ``D'') and 10 bidders (labeled 1--10). An arc
between a seller--bidder pair denotes an interaction, and the width
of the arc is proportional to the number of repeat-interactions
between the pair. Consider bidder 1 who has a total of 10
interactions, all of which are with seller A; we can say that bidder
1 is 100\% loyal to seller A. This is similar for bidders 2 and 3,
who have a total of 8 and 6 interactions, respectively, all of which
are, again, with seller A. In contrast, bidders~4 and 5 are only
80\% and 70\% loyal to seller A since, out of their total number of
interactions (both 10), they share 2 with seller B and 3 with seller
C, respectively. All-in-all, seller A attracts mostly highly loyal
bidders. This is different for seller D who attracts mostly little
loyal bidders, as he shares all of his bidders with either seller B
or C.

For each seller, we can summarize the proportion of loyal bidders
and the degree of their loyalty in the associated \textit{loyalty
distribution}. The loyalty distributions for sellers A--D are
displayed in the right panel of Figure \ref{fig:illustration}. The
$x$-axis denotes the degree of loyalty (e.g., 100\% or 80\%
loyal), and the $y$-axis denotes the corresponding density. We can see
that the shape of all four distributions is very different; while
seller A's distribution is very left-skewed (mostly high-loyal
bidders), seller D's distribution is very right-skewed (mostly
little-loyal bidders). The distributions of sellers B and C fall
somewhat in between, yet they are still very distinct from one
another.

Note that our definition of loyalty is similar to the concept of in-
and out-degree analysis. More precisely, we first measure the
proportion of interactions for each buyer (i.e., the normalized
distribution of out-degree). Then, we measure the perceived loyalty
of each seller, which can be viewed as the distribution of the
weighted in-degree. This definition of loyalty is very similar to
the concept of brand-switching in marketing. In essence, if we have
a fixed number of brands (sellers in our case) and a pool of buyers
(i.e., bidders), then we measure the switching-behavior from one
brand to another.

While the loyalty distributions in Figure \ref{fig:illustration}
capture all of the relevant information, we cannot use them for
further analysis (especially modeling). Thus, our next step is to
characterize each loyalty distribution by only a few numbers. To
that end, we employ a very flexible dimension reduction approach via
functional data analysis.

\subsection{From loyalty distributions to loyalty measures}

In order to investigate the effect of loyalty on the outcome of an
auction, we first need to characterize a seller's loyalty
distribution. While one could characterize the distributions via
summary statistics (e.g., mean, median or mode), Figure
\ref{fig:illustration} suggests that loyalty is too heterogeneous
and too dispersed. Therefore, we resort to a very flexible
alternative via functional data analysis [\cite{ramsay2005fda}].

By functional data we mean a collection of continuous functional
objects such as curves, shapes or images. Examples include
measurements of individuals' behavior over time, digitized 2- or
3-dimensional images of the brain, or recordings of 3- or even
4-dimensional movements of objects traveling through space and time.
In our application, we regard each seller's loyalty distribution as
a functional observation. We capture similarities (and differences)
across distributions via \textit{functional principal component
analysis} (fPCA), a functional version of principal component
analysis [see \cite{keiputikal2001}]. In fact, while
\citet{keiputikal2001} operate on the true probability
distributions, these are not known in our case; hence, we apply fPCA
to the observed (empirical) distribution function, which may
introduce an extra level of estimation error.

Functional principal component analysis is similar in nature to
ordinary PCA; however, rather than operating on data-vectors, it
operates on functional objects. In our context, we take the observed
loyalty distributions [i.e., the histograms
Figure~\ref{fig:illustration}(b)] as input. While one could also
first smooth the observed histograms, we decided against it since
the results were not substantially different.

Ordinary PCA operates on a set of data-vectors, say,
$\bx_1,\dots,\bx_n$, where each observation is a $p$-dimensional
data-vector $\bx_i = (x_{i1},\dots,x_{ip})^\mathrm{T}$. The goal of ordinary
PCA is to find a projection of $\bx_1,\dots,\bx_n$ into a new space
which maximizes the variance along each component of the new space
and at the same time renders the individual components of the new
space orthogonal to one another. In other words, the goal of
ordinary PCA is to find a PC vector $\be_1 =
(e_{11},\dots,e_{1p})^\mathrm{T}$ for which the principal component scores
(PCS)
%
\begin{equation}\label{eq: pc score}
S_{i1}= \sum_{j} e_{1j}x_{ij} = \be_1^\mathrm{T}\bx_i
\end{equation}
maximize $\sum_i S_{i1}^2$ subject to
%
\begin{equation}\label{eq: pc norm}
\sum_j e_{1j}^2 = \| \be_1 \|^2 = 1.
\end{equation}
This yields the first PC, $\be_1$. In the next step we compute the
second PC, $\be_2 = (e_{21},\dots,e_{2p})^\mathrm{T}$, for which, similarly
to above, the principal component scores $S_{i2}= \be_2^\mathrm{T}\bx_i$
maximize $\sum_i S_{i2}^2$ subject to $\| \be_2 \|^2 = 1$ and the
\textit{additional constraint}
%
\begin{equation}\label{eq: pc contraint}
\sum_j e_{2j}e_{1j} = \be_2^\mathrm{T} \be_1 = 0.
\end{equation}
This second constraint ensures that the resulting principal
components are orthogonal. This process is repeated for the
remaining PC, $\be_3,\dots,\be_p$.

The functional version of PCA is similar in nature, except that we
now operate on a set of continuous curves rather than discrete
vectors. As a consequence, summation is replaced by integration.
More specifically, assume that we have a set of curves
$\bx_1(s),\dots,\bx_n(s)$, each measured on a continuous scale
indexed by $s$. The goal is now to find a corresponding set of PC
curves, $\be_i(s)$, that, as previously, maximize the variance along
each component and are orthogonal to one another. In other words, we
first find the PC function, $\be_1(s)$, whose PCS
%
\begin{equation}\label{eq: fpc score}
S_{i1}= \int e_{1}(s)x_{i}(s)\, ds
\end{equation}
maximize $\sum_i S_{i1}^2$ subject to
%
\begin{equation}\label{eq: fpc norm}
\int e_{1}^2 \,ds = \| \be_1 \|^2 = 1.
\end{equation}
Similarly to the discrete case, the next step involves finding
$\be_2$ for which the PCS $S_{i2}= \int e_{2}(s)x_{i}(s)\, ds$
maximize $\sum_i S_{i2}^2$ subject to $\| \be_2 \|^2 = 1$ and the
additional constraint
%
\begin{equation}\label{eq: fpc contraint}
\int e_{2}(s)e_{1}(s)\, ds = 0.
\end{equation}

In practice, the integrals in (\ref{eq: fpc score})--(\ref{eq: fpc
contraint}) are approximated either by sampling the predictors,
$\bx_i(s)$, on a fine grid or, alternatively, by finding a lower-dimensional expression for the PC functions $\be_i(s)$ with the help
of a basis expansion. For instance, let
$\bphi(s)=(\phi_1(s),\dots,\phi_K(s))$ be a suitable basis expansion
[\cite{ramsay2005fda}], then we can write
%
\begin{equation}\label{eq:basis exp}
\be_i(s) = \sum_{k=1}^{K} b_{ik}\phi_k(s) = \bphi(s)^\mathrm{T} \bb_i
\end{equation}
for a set of basis coefficients $\bb= (b_{i1},\dots,b_{iK})$. In
that fashion, the integral in, for example, (\ref{eq: fpc contraint}) becomes
%
\begin{equation}\label{eq: fpc new contraint}
\int e_{2}(s)e_{1}(s)\, ds = \bb_{1}^\mathrm{T} \bW\bb_{2},
\end{equation}
where $\bW= \int\bphi(s)\bphi(s)^\mathrm{T}\, ds$. For more details, see
\citet{ramsay2005fda}. In this work we use the grid-approach.

\begin{figure}[b]

\includegraphics{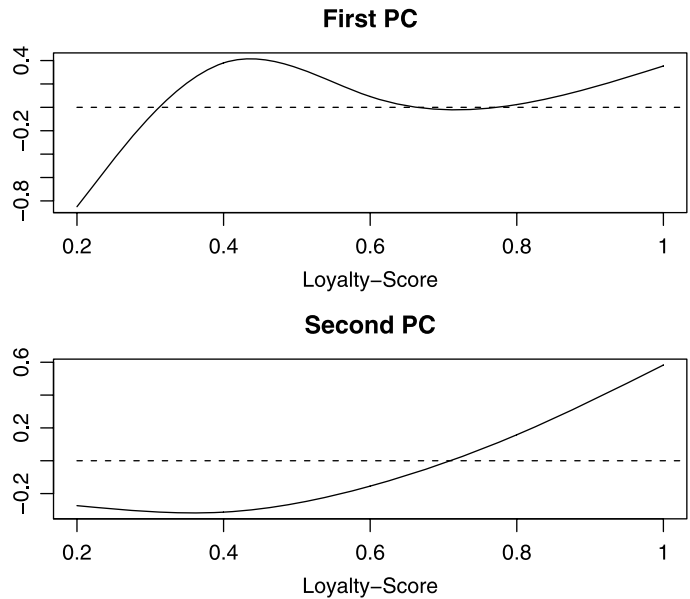}

\caption{First two principal component curves. The dashed line
indicates the $x$-origin.}
\label{fig:pc loadings}
\end{figure}

Common practice is to choose only those eigenvectors that correspond
to the largest eigenvalues, that is, those that explain most of the
variation in $\bx_1(s),\dots,\bx_n(s)$. By discarding those
eigenvectors that explain no or only a very small proportion of the
variation, we capture the most important characteristics of the
observed data patterns without much loss of information. In our
context, the first 2 eigenvectors capture over 82\% of the variation
in loyalty distributions.

\subsection{Interpreting the loyalty measures}

Since our loyalty measures are based on their principal component
representations, interpretation has to be done with care. Figure
\ref{fig:pc loadings} shows the first 2 principal components (PCs).
The first PC (top panel) shows a growing trend and, in particular,
it puts large negative weight on the lowest loyalty scores (between
0 and 0.2) while putting positive weight on medium to high loyalty
scores (0.4 and higher). Thus, we can say that the first PC contrasts
the extremely disloyal distributions from the rest. Table~\ref{pc_summaries} (first row) confirms this notion: Notice the large
negative correlation with the minimum; also, the large correlation
with the skewness indicates that PC1 truly captures extremes in the
loyalty distributions' scores and shape. We can conclude that PC1
distinguishes distributions of ``pure disloyalty'' from the rest.

The second PC has a different shape. The second PC puts most
(positive) weight on the highest loyalty scores (between 0.8 and 1);
it puts negative weight on scores at the medium and low scores
(between 0.4 and 0.6) and thus contrasts average loyalty from
extremely high loyalty. Indeed, Table~\ref{pc_summaries} (second
row) shows that PC2 has a high positive correlation with the maximum
and a high negative correlation with the median. In that sense, it
distinguishes the mediocre loyalty from the stars.

While the above interpretations help our understanding of the
loyalty components, their overall impact is still hard to grasp,
especially because every individual loyalty distribution will---by
nature of the principal component decomposition---comprise of a
different \textit{mix} between PC1 and PC2. Moreover, as we apply fPCA
to observed densities (i.e.,~histograms), individual values of each
density function must be heavily correlated. This adds additional
constraints on the PCs and their interpretations. Hence, in the
following, we discuss five \textit{theoretical} loyalty distributions
and their corresponding representation via PC1 and PC2.

\begin{table}
\tablewidth=210pt
\caption{Correlation between the first two PC scores and summary
statistics of sellers' loyalty distributions}\label{pc_summaries}
\begin{tabular*}{210pt}{@{\extracolsep{\fill}}ld{2.2}d{2.2}cd{2.2}c@{}}
\hline
& \multicolumn{1}{c}{\textbf{Median}} & \multicolumn{1}{c}{\textbf{SDev}} & \multicolumn{1}{c}{\textbf{Max}}
& \multicolumn{1}{c}{\textbf{Min}} & \multicolumn{1}{c@{}}{\textbf{Skew}} \\
\hline
PC1 & 0.55 & -0.2 & 0.52 & -0.99 & 0.77\\
PC2 & -0.78 & -0.05 & 0.81 & -0.02 & 0.63\\
\hline
\end{tabular*}
\end{table}

\begin{figure}[b]

\includegraphics{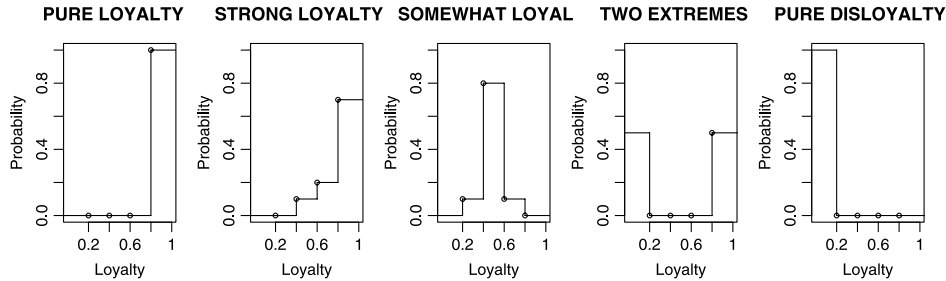}

\caption{Five theoretic loyalty distributions.}
\label{fig:theoretic loyalty}
\end{figure}

Take a look at Figure \ref{fig:theoretic loyalty}. It shows five
plausible loyalty distributions as they may develop out of a
bidder--seller network. We refer to these distributions as
``theoretic loyalty distributions'' and we can characterize them by
their specific shapes. For instance, the first distribution
is comprised of 100\% loyal buyers and we hence refer to it as ``pure
loyalty;'' in contrast, the last distribution is comprised of 100\%
disloyal buyers and we hence name this distribution ``pure
disloyalty;'' the distribution in the center (``somewhat loyal'') is
interesting since it is comprised mostly of buyers that exhibit some
loyalty but do not purchase exclusively from only one seller.

Table \ref{theoretical_loyalties} shows the corresponding PC scores.
We can see that the theoretical distribution corresponding to \textit{pure loyalty} scores very high on PC1 since it is very right-skewed
and does not have any values lower than 0.9; in contrast, notice the
PC1 scores for \textit{pure disloyalty}: while it is the mirror image
of \textit{pure loyalty}, it scores (in absolute terms) higher than the
former because it is not only very (left-) skewed, but its extremely
small values weigh heavily (and negatively) with the first part of
the PC1-shape, which is in contrast to the positive values of \textit{pure loyalty} which do not receive as much weight. As for PC2, Table
\ref{theoretical_loyalties} shows that \textit{pure loyalty} scores
even higher on that component as its values are extremely large,
much larger than the typical (median) loyalty values. In contrast,
\textit{pure disloyalty} has very small PC2 values, as low scores are
given very little weight by PC2.

\begin{table}
\caption{PC1 and PC2 scores for the theoretical loyalty
distributions from Figure \protect\ref{fig:theoretic loyalty}}
\label{theoretical_loyalties}
\begin{tabular}{@{}lccd{2.2}d{2.2}d{2.2}@{}}
\hline
& \textbf{Pure lyty} & \textbf{Strong lyty} & \multicolumn{1}{c}{\textbf{Somewhat loyal}}
& \multicolumn{1}{c}{\textbf{Two extremes}} & \multicolumn{1}{c@{}}{\textbf{Pure
dislyty}}\\ \hline
PC1 & 0.56 &0.47 &0.32 &-0.04 &-0.64\\
PC2 & 0.72 &0.51 &-0.08 &0.35 &-0.01\\
\hline
\end{tabular}
\end{table}

We can make similar observations for the remaining theoretical
loyalty distributions. For instance, the distribution of \textit{somewhat loyal} scores high on PC1 since it does not have many low values;
but it also only receives an average score on PC2 since it does not
have many high values either. In the following, we will use these
theoretical loyalty distributions to shed more light on the
relationship between loyalty and the outcome of an auction.

\section{Modeling e-loyalty}\label{sec4}

Our goal is to investigate the effect of loyalty on the outcome of
an auction. For instance, we would like to see whether sellers who
attract exclusively high-loyal bidders elicit price-premiums, or
whether more variability in buyers' loyalty leads to a higher price.
To that end, we start out, in similar fashion to many previous
studies on online auctions [e.g., \cite{Reiley00}], with an
ordinary least squares (OLS) modeling framework. That is, we
investigate a model of the form
%
\begin{equation}\label{eq: regression model}
{\mathbf y} = {\mathbf X} {\bolds\beta} + {\bolds\epsilon},
\end{equation}
where ${\mathbf X}$ is a matrix of covariates and ${\bolds
\epsilon}$ follows the standard linear model assumptions. For the
choice of the covariates, we are primarily interested in the effect
of loyalty on the price of an auction (i.e., the first 2 PC scores
from the previous section are our main interest). However, we also
want to control for factors other than loyalty which are also known
to have an impact on price; these factors include auction
characteristics (auction duration), item characteristics (item
quantity, size and pieces), seller characteristics (seller feedback,
i.e., reputation and seller volume) and auction competition (number
of bids, i.e., bid-count).

We first investigate a standard OLS model that relates these
covariates to price. However, we will show that an OLS approach
leads to violations of the model assumptions. The reason lies in the
asymmetry of the bidder--seller network: the presence of high-volume
sellers (i.e., seller nodes with extremely high degree) biases the
analysis and leads to wrong conclusions. In particular, high-volume
sellers have many repeat interactions which result in a strong
clustering of the data and thus violate the i.i.d. assumption
of OLS. We investigate several remedies to this problem. First, we
investigate two ``standard'' remedies via random effect (RE) models
and weighted least squares (WLS). Our results show that although
both remedies ease the problem, none removes it completely. We thus
argue that the data is too heterogeneous to be modeled within a
single model and compare our results with that of a
data-segmentation strategy.

\subsection{An initial model: OLS}

Many studies employ an OLS modeling framework to investigate
phenomena in online auctions such as the effect of the auction
format, the impact of a seller's reputation, or the amount of
competition [e.g., \cite{Reiley00}; \cite{bapavlou2002}; \cite{bapnaisr2008}]. However, one problem with an OLS model approach is
the presence of repeat observations on the same item. For instance,
if we want to study the effect of a seller's reputation (measured by
her feedback score), then repeat auctions by the same seller will
severely overweight the effect of high-volume sellers in the OLS
model. This problem is typically not addressed in the online auction
literature. We face a very similar problem when modeling the effect
of e-loyalty.

\begin{table}
\caption{Regression models on the entire (unsegmented) data set. The
top panel shows the results of OLS regression; the middle panel
shows the random effects model, and the bottom panel shows the
weighted least squares model. The response is always log-price}
\label{effect of theoretical loyalties}
\begin{tabular*}{\textwidth}{@{\extracolsep{\fill}}ld{2.2}cc@{}}
\hline
\textbf{Coefficient} & \multicolumn{1}{c}{\textbf{Estimate}} & \textbf{Std. error} &
\textbf{$\bolds p$-value} \\
\hline
(Intercept) & -1.64 & 0.06 & 0.00 \\
Auction duration & 0.02 & 0.00 & 0.00 \\
log(item quantity $+$ 1) & -0.04 & 0.04 & 0.30 \\
Bid count & 0.06 & 0.00 & 0.00 \\
log(Pieces) & 0.42 & 0.00 & 0.00 \\
Size & 0.07 & 0.00 & 0.00 \\
log(seller feedback $+$ 1) & -0.00 & 0.01 & 0.52 \\
Loyalty-PC1 & -0.17 & 0.05 & 0.00 \\
Loyalty-PC2 & -1.00 & 0.07 & 0.00 \\
log(Volume) & 0.16 & 0.01 & 0.00 \\
[3pt]
{AIC} & \multicolumn{1}{c}{15,148}\\
{$ R$-squared} & 0.77\\
[6pt]
(Intercept) & -0.58 & 0.22 & 0.00 \\
Auction duration & 0.01 & 0.00 & 0.00 \\
log(item quantity $+$ 1) & -0.21 & 0.07 & 0.00 \\
Bid count & 0.05 & 0.00 & 0.00 \\
log(Pieces) & 0.27 & 0.00 & 0.00 \\
Size & 0.03 & 0.00 & 0.00 \\
log(seller feedback $+$ 1) & 0.07 & 0.04 & 0.11 \\
Loyalty-PC1 & -0.40 & 0.22 & 0.07 \\
Loyalty-PC2 & -0.15 & 0.22 & 0.51 \\
log(Volume) & 0.05 & 0.04 & 0.20 \\
[3pt]
{AIC} & \multicolumn{1}{c}{8546} \\
{$ R$-squared} & \multicolumn{1}{c}{N/A}\\
[6pt]
(Intercept) & -1.59 & 0.12 & 0.00 \\
Auction duration & 0.01 & 0.00 & 0.01 \\
log(item quantity $+$ 1) & 0.30 & 0.10 & 0.00 \\
Bid count & 0.08 & 0.00 & 0.00 \\
log(Pieces) & 0.26 & 0.01 & 0.00 \\
Size & 0.00 & 0.00 & 0.61 \\
log(seller feedback $+$ 1) & -0.01 & 0.01 & 0.59 \\
Loyalty-PC1 & -1.00 & 0.08 & 0.00 \\
Loyalty-PC2 & 1.33 & 0.11 & 0.00 \\
log(Volume) & 0.24 & 0.01 & 0.00 \\
[3pt]
{AIC} &\multicolumn{1}{c}{55,073}\\
{$ R$-squared} & 0.43\\
\hline
\end{tabular*}
\end{table}

\begin{figure}

\includegraphics{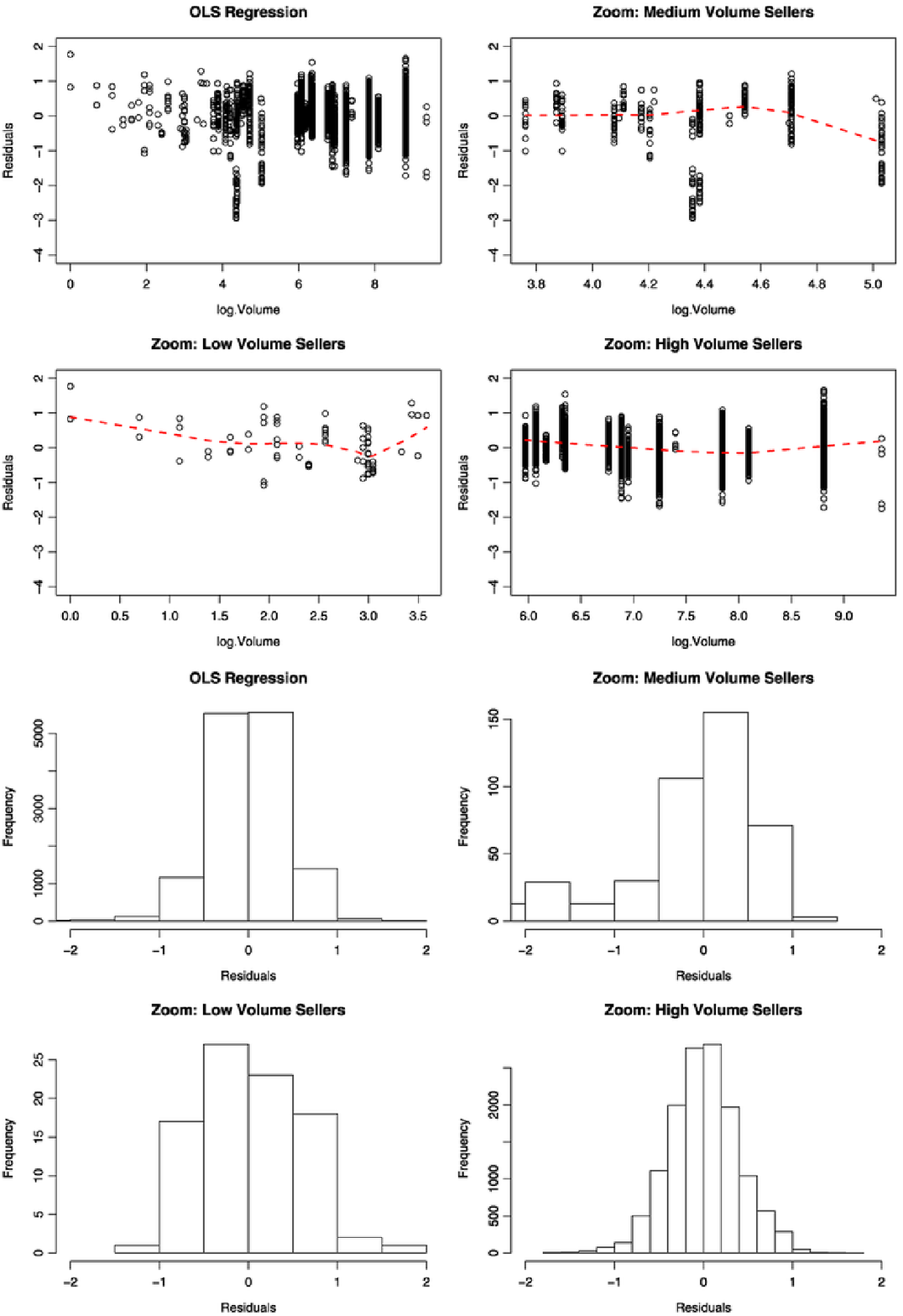}

\caption{Residuals of the OLS model.}
\label{fig:res2 ols}
\end{figure}

For illustration, take the OLS regression model in the top panel of
Table \ref{effect of theoretical loyalties}.
In this model we estimate the dependency of
(log-)price on loyalty (measured by PC1 and PC2), controlling for
all other factors described above. Note that this model appears to
fit the data very well ($R$-squared $=$ 77\%). However, it is curious to
see that seller feedback has a negative sign and is statistically
insignificant. This contradicts previous findings which found that
an increased level of trust leads to price premiums
[\cite{bajari2004eii}; \cite{bapavlou2002}; \cite{Reiley00}].

Figure \ref{fig:res2 ols} shows the residuals corresponding to the
above model. The top half shows the residuals plotted against
seller-volume; the bottom half shows the residual distribution. For
each type of graph we present 4 different views: one graph (left
graphs in first and third panel) gives an overview; the other graphs
zoom in by seller volume (low, medium and high volume,
respectively). Notice that the residuals are rather skewed: a large
proportion of residuals are negative (see, e.g., top left graph),
implying that our model over-estimates price effects of loyalty.
Moreover, we can also see that the residual-variation increases for
larger seller volumes. If we zoom in on both the low-volume and
medium-volume sellers, we can see that the true effect of model
misspecification is confounded with seller volume: while price
effects of low-volume sellers are underestimated (note the
positive-skew in the residual distribution for low-volume sellers),
the effects are overestimated for medium-volume sellers
(negative-skew); only high volume sellers appear to be captured well
by the model. Thus, the OLS regression model blends low volume and
medium volume sellers but represents neither of them adequately.

\subsection{Two alternate models: WLS and RE}

We have seen in the previous section that an OLS approach does not
result in a model that can be interpreted without concerns. We thus
investigate two alternate models, a random effects (RE) model and
weighted least squares (WLS).

\begin{figure}

\includegraphics{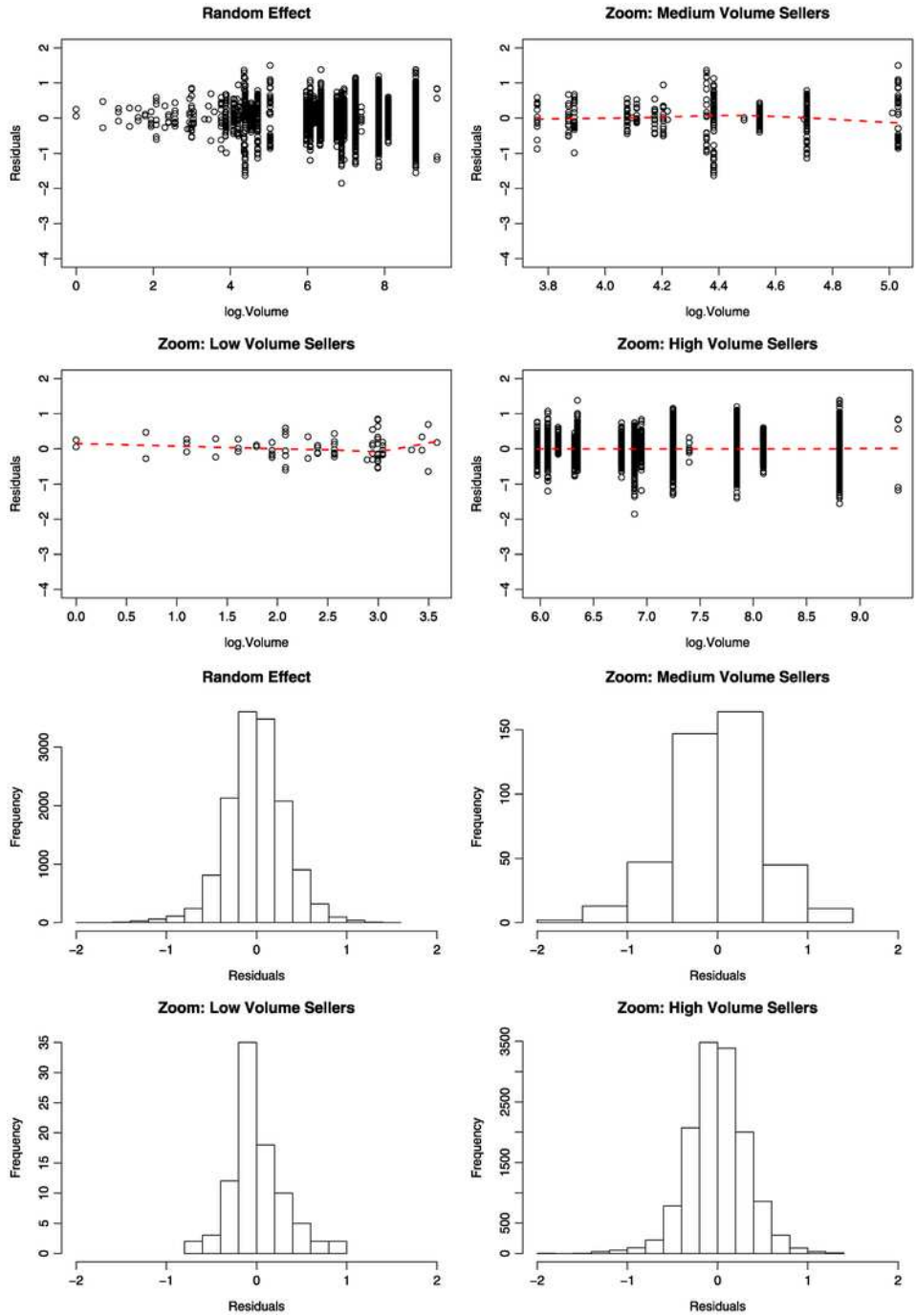}

\caption{Residuals of the RE model.}
\label{fig:res2 random}
\end{figure}

Random effects models are often employed when there are repeat
observations on the same subject or when the data is clustered
[e.g., \cite{agresti2000}]. Since we have many repeat auctions by
the same seller, adding a random, seller-specific effect to the
model in (\ref{eq: regression model}) lends itself as a natural
remedy for OLS. While RE models have become popular only recently
with the advent of powerful computing and efficient
algorithms,\footnote{Quite often, RE models have to be estimated
using computationally intensive techniques such as MCMC or other
forms of stochastic estimation.} WLS has been around for a longer
time as a possible solution for heteroscedasticity
[\cite{greene2003}]. While the principle of WLS is powerful, it
assumes that the matrix of weights is known (or at least known up to
a parameter value), which reduces its practical value. In our
context, we use weights that are inversely proportional to the
residual variance in each cluster. We will now compare both
approaches and see if they result in more plausible models for
e-loyalty.\looseness=-1

Table \ref{effect of theoretical loyalties} (second and third panels) shows the results of the
RE and WLS models, respectively. We can see that WLS results in a
very poor model fit (both in terms of $R$-squared and AIC). While the
RE model results in much better model fit (compared to both the WLS
and the OLS model), it is curious that seller feedback is
insignificant, similar to the OLS model above. In fact, it is quite
curious that none of the seller-related variables (feedback, loyalty
or volume) are significant in the RE model. This finding suggests
that none of the actions taken by the seller affect the outcome of
an auction, which contradicts both common practitioner knowledge as
well as previous research on the topic [\cite{bajari2004eii}; \cite{bapavlou2002}; \cite{Reiley00}].

Figure \ref{fig:res2 random} shows the residuals of the RE model. We
can see that the magnitude of the residuals has decreased,
suggesting a better model fit. This is expected as the random
effects account for seller-specific variation due to individual
selling strategies (e.g., seller-specific auction parameters or
product descriptions), which all may lead to differences in final
price. But we can also see that the RE model still suffers from
heteroscedasticity (much larger residual variance for high volume
sellers compared to low volume sellers).

\begin{figure}

\includegraphics{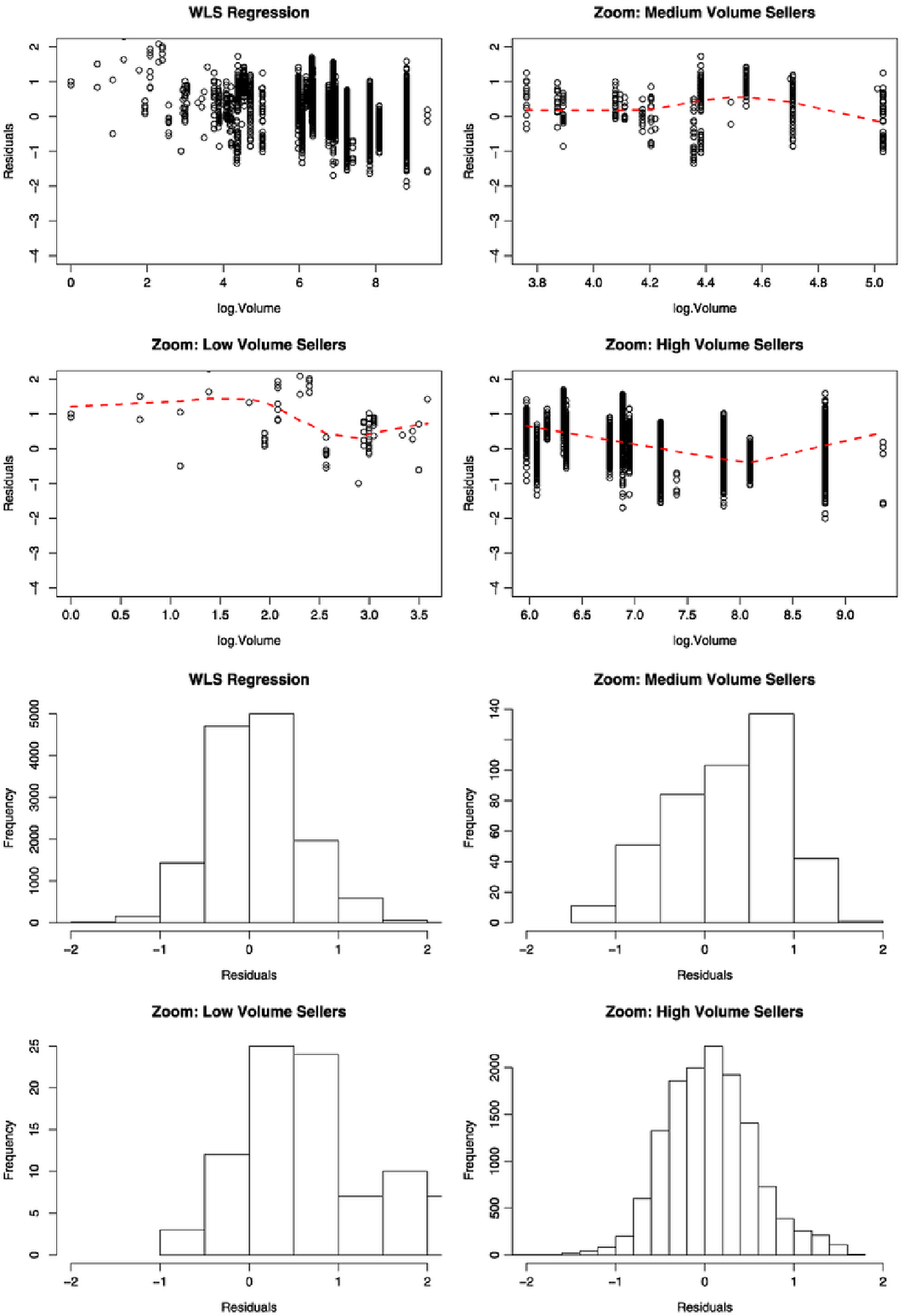}

\caption{Residuals of the WLS model.}
\label{fig:res2 wls}
\end{figure}

Figure \ref{fig:res2 wls} shows the corresponding residuals of the
WLS approach. While we would have expected that WLS tames the
heteroscedasticity somewhat, it appears that model fit has become
worse. (This is also supported by the much poorer values of
$R$-squared and AIC.) One possible reason is that weights have to be
chosen by the user (inversely proportional to seller volume, in our
case), which may not result in the most appropriate weighting of the
data.

\subsection{Data segmentation}

None of the proposed modeling alternatives so far have lead to
models with reasonable residuals or economically defendable
conclusions. In fact, we have seen that the model fit differs
systematically by the seller volume. We take this as evidence that
the data may be segmented into different \textit{seller volume
clusters}. We have seen earlier (e.g., Figures \ref{fig:net 2} and
\ref{fig:net 4}) that sellers of different magnitude exhibit quite
different effects on bidders. We will thus now first cluster the
data and then model each data-segment separately.

We first cluster the data by seller volume (low, medium and high)
and then apply OLS regression within each segment, resulting in
three different regression models, one for each segment. We select
the clusters with the objective of minimizing the residuals mean
squared errors within each cluster. This results in the following
three segments: Low volume sellers---40 transactions or less;
medium volume sellers---40--350 transactions; high volume sellers---more than 350 transactions.

\begin{figure}

\includegraphics{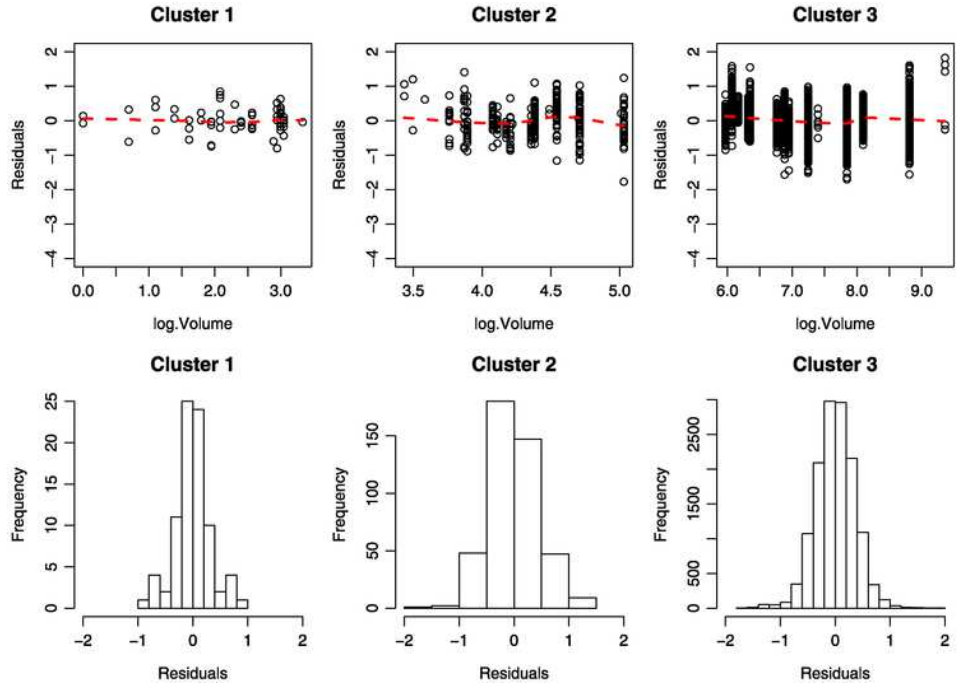}

\caption{Residuals after data segmentation. The left panel shows
the residuals of an OLS model fitted only to the data from cluster 1 (low
volume sellers); the middle and right panels show the corresponding
residuals for medium and high volume sellers.}
\label{fig:res cluster}
\end{figure}

Figure \ref{fig:res cluster} shows the residuals of the resulting
three models. We can see that the model fit is much better compared
to the previous modeling approaches. In each segment the magnitude
of the residuals is very small, all residuals scatter around the
origin, and we also no longer find evidence for heteroscedasticity
in any of the three segments. In fact, the model fit statistics (see
Table \ref{reg:cluster}) suggest that the segmentation approach
leads to a much better representation of the data compared to either
OLS, RE or WLS models.

Table \ref{reg:cluster} shows the parameter estimates for each
segment. We can see that the relationship between loyalty, trust and
price varies from segment to segment. In fact, while for the low
volume sellers the significance of all seller-related variables
(feedback, loyalty or volume) is low, both feedback and volume are
much more significant than loyalty. (Note the much smaller $p$-values
of seller feedback and volume.) This suggests that while
seller-related actions may not play much of a role for low volume
sellers (such as rookie sellers and sellers that are new to the
market), trust is much more important compared to loyalty. This
makes sense as low volume sellers have not much of a chance to
establish a loyal customer base due to the infrequency of their
transactions.

This is different for medium volume sellers. For medium volume
sellers, loyalty and volume are more significant than feedback. This
suggests that with increasing frequency of transactions, repeat
transactions (i.e., loyalty) have a more dominant effect on a
seller's bottom line. This effect is even more pronounced for high
volume sellers. This suggests that high volume sellers are most
affected by the actions of repeat customers. It is also interesting
that both feedback and loyalty are significant for high volume
sellers. This suggests that in the presence of two sellers with the
same reputation, buyers ``act with memory'' and return to repeat
their previous shopping experience.

In order to precisely quantify the effect of loyalty in each
segment, consider Table \ref{tblOLS}. In
that table, we present, for each of the 5 theoretical loyalty
distributions from Figure \ref{fig:theoretic loyalty}, their
corresponding combined effect on the regression model. That is, we
compute the combined effect of PC1 and PC2, holding all other
variables in the model constant.

\begin{table}
\tablewidth=300pt
\caption{Regression models on three data segments. The top panel
shows the result for the low volume sellers; middle panel shows
medium volume sellers; bottom panel shows high volume sellers. The
response is always log-price} \label{reg:cluster}
\begin{tabular*}{300pt}{@{\extracolsep{\fill}}ld{2.2}cc@{}}
\hline
\textbf{Coefficient} & \multicolumn{1}{c}{\textbf{Estimate}} & \textbf{Std. error} &
\textbf{$\bolds p$-value} \\
\hline
(Intercept) & -0.51 & 0.36 & 0.16 \\
Auction duration & 0.04 & 0.02 & 0.10 \\
log(item quantity $+$ 1) & -0.06 & 0.10 & 0.50 \\
Bid count & 0.12 & 0.01 & 0.00 \\
log(Pieces) & 0.15 & 0.06 & 0.02 \\
Size & 0.06 & 0.03 & 0.05 \\
log(seller feedback $+$ 1) & 0.08 & 0.04 & 0.07 \\
Loyalty-PC1 & -0.22 & 0.16 & 0.18 \\
Loyalty-PC2 & -0.03 & 0.13 & 0.82 \\
log(Volume) & -0.10 & 0.06 & 0.07 \\
[3pt]
 {AIC} & \multicolumn{1}{c}{63} & & \\
 {$ R$-squared} & 0.76 & & \\
[6pt]
(Intercept) & -0.49 & 0.38 & 0.20 \\
Auction duration & -0.05 & 0.01 & 0.00 \\
log(item quantity $+$ 1) & -0.11 & 0.19 & 0.58 \\
Bid count & 0.17 & 0.01 & 0.00 \\
log(Pieces) & 0.03 & 0.01 & 0.05 \\
Size & 0.01 & 0.01 & 0.58 \\
log(seller feedback $+$ 1) & -0.01 & 0.02 & 0.65 \\
Loyalty-PC1 & -0.24 & 0.12 & 0.04 \\
Loyalty-PC2 & 0.28 & 0.19 & 0.15 \\
log(Volume) & 0.33 & 0.08 & 0.00 \\
[3pt]
{AIC} & \multicolumn{1}{c}{534}\\
{$ R$-squared} & 0.75\\
[6pt]
(Intercept) & 1.68 & 0.10 & 0.00 \\
Auction duration & 0.02 & 0.00 & 0.00 \\
log(item quantity $+$ 1) & -0.25 & 0.04 & 0.00 \\
Bid count & 0.05 & 0.00 & 0.00 \\
log(Pieces) & 0.39 & 0.00 & 0.00 \\
Size & 0.06 & 0.00 & 0.00 \\
log(seller feedback $+$ 1) & 0.42 & 0.01 & 0.00 \\
Loyalty-PC1 & 7.39 & 0.16 & 0.00 \\
Loyalty-PC2 & -9.41 & 0.17 & 0.00 \\
log(Volume) & -0.78 & 0.02 & 0.00 \\
[3pt]
{AIC} & \multicolumn{1}{c}{10,482} \\
{$ R$-squared} & 0.83\\
\hline
\end{tabular*}
\end{table}

\begin{table}
\tabcolsep=0pt
\caption{Quantifying the effect of the 5
theoretic loyalty distributions from Figure \protect\ref{fig:theoretic loyalty}}
\label{tblOLS}
\begin{tabular*}{\textwidth}{@{\extracolsep{\fill}}ld{2.2}d{2.2}d{2.2}d{2.2}d{2.2}@{}}
\hline
& \multicolumn{1}{c}{\textbf{Pure lyty}} & \multicolumn{1}{c}{\textbf{Strong lyty}}
& \multicolumn{1}{c}{\textbf{Somewhat loyal}} & \multicolumn{1}{c}{\textbf{Two extremes}}
& \multicolumn{1}{c@{}}{\textbf{Pure dislyty}}\\
\hline
Cluster 1 (low) &-0.14 &-0.12 &-0.07 &0.00 &0.14\\
Cluster 2 (medium) & 0.07 &0.03 &-0.10 &0.11 &0.15\\
Cluster 3 (high) &-2.60 &-1.29 &3.14 & -3.61 &-4.62\\
\hline
\end{tabular*}
\end{table}

We can see that in clusters 1 and 2, the effect of loyalty is
considerably small, consistent with the small (and insignificant)
coefficients for low and medium volume sellers in Table
\ref{reg:cluster}. For cluster 3, it is interesting that only the
distribution corresponding to \textit{somewhat loyal} buyers results in
a positive price effect. In fact, we can see that extreme loyalty
(i.e., the distributions for both \textit{pure loyalty} and \textit{pure
disloyalty}) has negative implications for a seller's bottom line.
While the effect of disloyal bidders is easier to explain (disloyal
bidders may ``shop around'' more actively in the search for lower
prices and, as a result, drive down a seller's revenue), the negative
effect of purely loyal bidders may be due to the fact that a bidder
who exclusively interacts with the same seller may form an opinion
about that seller's ``going price'' which results in a less
competitive auction process (and thus renders the transaction into a
fixed-price transaction). Thus, our results show that the effect of
loyalty is surprisingly ``nonlinear'' in that a mix of somewhat loyal
bidders results in the most competitive auction environment and thus
the highest price for the seller.

Another way of quantifying the impact of loyalty is via the
difference between pure loyalty and pure disloyalty. Notice that the
difference in estimated coefficients equals ($-2.60-4.62)\approx$ 2,
which (as the response is on the log-scale) implies that, all else
equal, sellers with a purely loyal customer base extract price
premiums 200\% higher compared to sellers with a purely disloyal
customer base.

\section{Conclusion}\label{sec5}

In this paper we investigate loyalty of online transactions. Loyalty
is an important element to many business models, and it is
especially difficult to manage in the online domain where consumers
are offered different choices that are often only a mouse-click
away. We study loyalty in online auctions. We derive online loyalty
from the network of sellers and bidders and find that while bidder's
loyalty can have a strong impact on the outcome of an auction, the
magnitude of its impact varies depending on the size of the seller.

We want to point out that while we find that loyalty has a strong
effect on price, we do not determine the cause of loyalty. A buyer's
loyalty can have many different causes such as a high-quality
product, a speedy delivery, or an otherwise seamless service. While
loyalty could also be caused by price itself (i.e., a buyer returning
to the same seller because of a low price), it is unlikely in our
setting due to the auction process. Recall that in an auction the
price is not fixed. Thus, a seller offering a top notch product and
an outstanding service will sooner or later see an increase in
bidders and, as a result, more competition and thus a higher price
for her product. Thus, loyalty is unlikely to be caused merely by
bargain sellers.\looseness=1

Also, we want to emphasize that while we find many repeat
transactions between the same seller--bidder pair in our data, the
frequency of these repeat interactions may depend on the type of
product and the buyer's demand for this product. In our case (beads,
i.e., arts and crafts), buyers have frequently re-occurring demand
for the same product and, hence, the chances that a buyer will
seek out the same seller rise drastically. On the other hand, if we
were to consider the market for a product in which repeat
transactions are less common (such as computers, digital cameras,
automobiles, etc.), our loyalty networks would likely not be as
dense. Nevertheless, it would be equally important for sellers to
understand what factors drive consumers to spend money and we
believe that loyalty networks are one way to address that question.

There are several statistical challenges when studying loyalty
networks. First, deriving quality measures from the observed
networks requires a method that can capture both the intensity as
well as the size of loyalty. We accomplish this using ideas from
functional data analysis. Second, modeling the effect of loyalty is
complicated by the extreme skew of loyalty networks. Our analysis
shows that many different approaches can lead to model
misspecification and, as a consequence, to economically wrong
conclusions. Similar problems likely exist in other studies on
online markets (e.g., those that study seller feedback or reputation
where one also records repeat observations on the same seller). Our
analysis leads us to conclude that the data is too segmented to be
treated by a single model and thus propose a data-clustering
approach.

Another statistical challenge revolves around \textit{sampling}
bidder--seller networks. As pointed out earlier, we have the complete
set of bidding records for a certain product (Swarovski beads, in
this case) for a certain period of time (6 months). As a result, we
have the complete bidder--seller network for this product, for this
time frame. While sampling would be an alternative, it would result
in an incomplete network (since we would no longer observe all
nodes/arcs). As a result, we would no longer be able to compute
loyalty without error, which would bring up an interesting
statistical problem. But we caution that sampling would have to be
done very carefully. While one could, at least in theory, sample
randomly across all different eBay categories, it would bring up
several problems. The biggest problem is that we would now be
attempting to compare loyalty across different product types. For
instance, we would be comparing, say, a bidder's loyalty for
purchasing beads (a very low price, low stake item) with that of
purchasing digital cameras, computers, or even automobiles (all of
which are high price and high stakes), which would be conceptually
very questionable.

We also want to mention that we treat the bidder--seller network as
\textit{static} over time. Our data spans a time-frame of only 6 months
and we assume that loyalty is static over this time-frame. This
assumption is not too unrealistic as many marketing models consider
loyalty to be static over much longer time frames
[\cite{faderhardie2006}; \cite{faderhardielee2006}; \cite{donkers2003}].
While incorporating a temporal dimension (e.g., by using a network
with a sliding window or via down-weighting older interactions)
would be an intriguing statistical challenge, it is not quite clear
how to choose the width of the window or the size of the weights.
Moreover, we also explicitly tested for learning effects by buyers
over time and could not find any strong statistical evidence for it.

And finally, in this work we address one specific kind of network
dependence, namely, that between buyers and sellers. We argue that
the lack of independence among observations on the same sellers
leads to a clustering-effect and we investigate several remedies to
this challenge. However, the dependence structure may in fact be far
more complex. As bidders are linked to sellers which, in turn, are
linked again to other bidders, the true dependence structure among
the observations may be far more complex. This may call for
innovative statistical methodology and we hope to have sparked some
new ideas with our work.

\printaddresses

\end{document}